\documentclass[12pt]{iopart}

\usepackage{graphicx}
\usepackage{amsfonts}
\usepackage{tikz}

\begin{document}

\title[Probing criticality in quantum spin chains with neural networks]{Probing criticality in quantum spin chains with neural networks}

\author{A Berezutskii$^{1,2,3}$, M Beketov$^3$, 
D Yudin$^3$, Z Zimbor\'as$^{4,5,6}$ and J~D~Biamonte$^3$}

\address{$^1$ 1QB Information Technologies (1QBit), Waterloo, ON, N2J 1S7, Canada}
\address{$^2$ Moscow Institute of Physics and Technology, Dolgoprudny, Moscow Region 141700, Russia}
\address{$^3$ Deep Quantum Laboratory, Skolkovo Institute of Science and Technology, Moscow 121205, Russia}
\address{$^4$ Wigner Research Centre for Physics, Theoretical Physics Department, Budapest H-1525, Hungary}
\address{$^5$ MTA-BME Lend\"ulet Quantum Information Theory Research Group, Budapest, Hungary}
\address{$^6$ Mathematical Institute, Budapest University of Technology and Economics, Budapest, Hungary}

%\ead{berezutskii@phystech.edu}

\begin{abstract}
The numerical emulation of quantum systems often requires an exponential number of degrees of freedom which translates to a computational bottleneck.  Methods of machine learning have been used in adjacent fields for effective feature extraction and dimensionality reduction of high-dimensional datasets. Recent studies have revealed that neural networks are further suitable for the determination of macroscopic phases of matter and associated phase transitions as well as efficient quantum state representation. In this work, we address quantum phase transitions in quantum spin chains, namely the transverse field Ising chain and the anisotropic XY chain, and show that even neural networks with no hidden layers can be effectively trained to distinguish between magnetically ordered and disordered phases.  Our neural network acts to predict the corresponding crossovers finite-size systems undergo. Our results extend to a wide class of interacting quantum many-body systems and illustrate the wide applicability of neural networks to many-body quantum physics.
\end{abstract}

%\submitto{Journal of Physics: Complexity}

\maketitle

\section{Introduction}

The concept of deep learning \cite{LeCun2015} has attracted dramatic interest over the last decade. First applied in the domain of image and natural speech recognition, algorithms for machine learning have recently shown their utility in statistical mechanics of interacting classical and quantum systems \cite{Zhang2019,Schindler2017,Deng2017,Zhang2018,Venderley2018, liu2018discriminative,Wang2016,Hu2017,Chng2017,Chng2018, Carleo2017,Glasser2018,Cai2018, carrasquilla2019reconstructing, hibat2020recurrent, beach2019qucumber}.

Solving a quantum many-body problem often implies a coarse-graining procedure to remove redundant degrees of freedom from the short-range, or the high-energy, sector of the theory. In this case, a proper elucidation of low energy properties of the system or the type of its long-range ordering encodes the macroscopic behavior. In its turn, the methodology of machine learning in multidimensional and typically non-structured datasets is inevitably linked to the effective approaches to dimensionality reduction, thereby yielding a powerful technique for the detailed analysis of classical and quantum models in many-body physics \cite{RevModPhys.91.045002, carrasquilla2020machine}. Practical application of neural networks in the context of both supervised and unsupervised machine learning has now become commonplace for testing thermal, quantum, and topological phase transitions \cite{Zhang2019,Schindler2017,Deng2017,Zhang2018,Venderley2018, liu2018discriminative,Wang2016,Hu2017,Chng2017,Chng2018, kharkov2020revealing} as well as for formulating effective variational wave function ans\"atze states \cite{Carleo2017,Glasser2018,Cai2018, carrasquilla2019reconstructing, hibat2020recurrent, beach2019qucumber, Szabo2020}. The application of machine learning to quantum-information problems has also received significant interest recently, promising to directly probe the entanglement entropy \cite{Torlai2017,Koch2018,Rocchetto2018} as well as other properties. The utility of machine learning methods for quantum information purposes is driven by its great success in condensed matter physics \cite{Arsenault2014,Torlai2016,Carrasquilla2017,Nieuwenburg2017,Wetzel2017,Saito2017,Zhang2018,Mills2018,Choo2018,Bukov2018,Liu2018,Shirinyan2019,Burzawa2019, westerhout2020generalization,2020arXiv200500544U} and computational many-body methods \cite{Huang2017,Nagai2017,Xia2018,Suwa2019,de2020reconstructing, inack2018projective,mcnaughton2020boosting}. In this study, we employ a specific machine learning technique to create a low-dimensional representation of microscopic states, relevant for macroscopic phase identification and probing phase transitions. More specifically, we explore phase transitions in the transverse field Ising- and the anisotropic XY chains and demonstrate that even the simplest possible neural network architecture --- a binary classifier as a perceptron with no hidden neurons present is capable of keeping track of its macroscopic phases depending on the, e.g., external magnetic field or anisotropy parameter, without any prior knowledge. It is worth mentioning that an approach to spin models in higher dimensions has recently appeared and based on exact calculations of entanglement \cite{Xu2010}.

\section{Model systems}

\subsection{Transverse field Ising model}

One-dimensional spin models represent strongly correlated quantum systems that can be rigorously approached at equilibrium \cite{Pikin1966}.  Certain non-equilibrium properties can also be extracted \cite{Brandt1977}. In the following, we focus on the one-dimensional ferromagnetic transverse field Ising model (TFIM).  The TFIM naturally appears upon solving a classical two-dimensional Ising model with ferromagnetic-type nearest-neighbor exchange coupling and its exact solution dates back to the original works~\cite{Katsura1962,schultz1964two,pfeuty1970one}. Generally, the TFIM of $L$ spins on a chain with open boundary conditions is specified by the following Hamiltonian:
\begin{equation}\label{eq:1}
H = -J \sum\limits_{i=1}^{L-1} \sigma_i^z \sigma_{i+1}^z - \tau\sum\limits_{i=1}^L \sigma_i^x,
\end{equation}
which represents a $2^L\times2^L$ matrix with
$\sigma^{\alpha}_i$ ($\alpha=x,y,z$)  being a Pauli matrix acting on site $i$, and $J$ and $\tau$ stand for the strength of exchange coupling and external magnetic field respectively. Interestingly, despite its relative simplicity, this model was used to describe intricate physics, e.g., the order-disorder transitions in ferroelectric crystals of KH$_2$PO$_4$. At zero temperature, quantum fluctuations may lead to a restructuring of the ground state which is manifested by a certain non-analyticity in the ground state energy of the quantum Hamiltonian. For the case of the  Hamiltonian \eref{eq:1}, when there is no magnetic field present ($\tau=0$) the ground state configuration is purely determined by the exchange interaction, the first term in \Eref{eq:1}, which favors collinear magnetic ordering. For $J>0$, the ferromagnetic state is energetically preferable, meaning that all magnetic moments point in the same direction $\langle \sigma^z_i \rangle=+1$ (or $-1$), signaling the double degeneracy of the ground state. Increasing the transverse field beyond the critical value $\tau=\tau_c$ makes the system susceptible to spin flip and all the spins aligned in $x$ direction in the limit $\tau \rightarrow \infty$, i.e., disordered in $\sigma_z$ basis.

The one-dimensional TFIM can be worked out analytically by virtue of the Jordan-Wigner transformation that maps an interacting spin model onto that of free spin-polarized fermions \cite{pfeuty1970one,Sachdev2011}. The exact solution unambiguously demonstrates a continuous quantum phase transition (QPT) upon passing through the critical field $\tau_c = 1$ (in the units of $J$), separating magnetically ordered ferromagnet ($\tau < \tau_c$) and disordered paramagnetic states ($\tau > \tau_c$). Although there is no exact analytical solution in higher dimensional systems, a quantum phase transition can be clearly detected \cite{Sachdev2011}. It is worth noting that the phase diagram of a one-dimensional TFIM is very similar to that of a two-dimensional classical Ising model at finite temperature with a temperature-driven phase transition. Interestingly, this dualism has a strict mathematical form corresponding to the so-called Suzuki-Trotter decomposition and which maps a $d$-dimensional quantum model to a $d+1$ dimensional classical one \cite{Suzuki1976}.

\subsection{Anisotropic XY model}

The XY model is yet another well-known quantum spin lattice model of magnetism. One can arrive to the isotropic version of this model by switching off the ZZ couplings in the Heisenberg Hamiltonian. In its turn, the anisotropic XY model is a generalization of it in the sense that the interaction strength in the XY plane is not isotropic anymore. In this study, we limit ourselves to the case when there is no field transverse to the interaction plane. The Hamiltonian of the model is thus given by
\begin{equation}\label{eq:2}
H = -J \sum\limits_{i=1}^{L-1} \left(\frac{1+\gamma}{2} \sigma_i^x \sigma_{i+1}^x + \frac{1-\gamma}{2} \sigma_i^y \sigma_{i+1}^y\right),
\end{equation}
where $\gamma$ is the anisotropy parameter that is usually restricted to $-1 \leq \gamma \leq 1$ and $J$ is the coupling strength which we set to $1$ hereafter. If one sets $\gamma=0$ the fully isotropic case, which possesses an additional symmetry $[H,\sigma_i^z]=0$, is restored. On the other hand, it is also well-known that in the opposite case, i.e.~$\gamma = 1$, the ground state possesses a long-range Neel order which yields
\begin{equation}
    \sigma_i^x | \sigma \rangle = (-1)^i | \sigma \rangle
\end{equation}
and
\begin{equation}
    \sigma_i^y | \sigma \rangle = (-1)^i | \sigma \rangle
\end{equation}
for $\gamma=-1$ accordingly, as is described in detail in Ref.~\cite{lieb1961two}. It is clear that as $\gamma$ decreases from 1 to $-1$, the $x$- and $y$-components begin to compete. Its phase diagram is thus given by an $x$- and $y$-ferromagnetic states for $\gamma=1$ and $-1$ accordingly. The model is fully isotropic at $\gamma=0$ and undergoes a second-order phase transition at this point while the gap continuously vanishes \cite{lieb1961two,luo2018fidelity}.

\section{Methodology}

\subsection{General overview}

The complexity of a generic quantum many-body problem grows exponentially with the size of a system (using the best known methods), making the available numerical routines computationally demanding. While machine learning has been specifically designed to coarse-grain certain information while maintaining  relevant and unique features corresponding to the dataset (reminiscent to the formalism of renormalization group in statistical and high-energy physics \cite{mehta2014exact}) it appears to be perfectly suited for identification of classical and quantum phases \cite{Carrasquilla2017,Tanaka2017,Morningstar2017}. Indeed, sampled spin-$\frac{1}{2}$ configurations can be mapped to either binary numbers or black and white pixels which can be further classified in the form of macroscopic configurations, representing the class of problems which machine learning has been routinely used for. However, typically for quantum many-body systems we do not have predefined labels, so the use of unsupervised learning is favored. Within this paradigm we search for clusterization or associative rules that govern the behavior of a system. Unsupervised learning can also take measurement data and essentially reconstruct the wave function from individual images or snapshots. These reconstruction techniques based on machine learning are now being studied and compared to traditional techniques based on quantum state and quantum process tomography \cite{Wang2016,Broecker2017,Huembeli2018,Nieuwenburg2017,macarone2019experimental}.

The advantage of using machine learning algorithms for exploration of both classical and quantum phase transitions is associated with finding certain features related to symmetry breaking in microscopic configurations. Particularly, phase transitions in magnetically ordered systems result in spin directions being randomized by the temperature---while the corresponding temperature can be detected as a point where the magnetization drops. When considering quantum phase transitions one typically investigates a finite region of sudden change that shrinks in the thermodynamic limit to a single point of non-analyticity \cite{Vojta2003}. Alternatively, in the vicinity of a phase transition point one can examine the behavior of the order parameter, which is known to collapse, or the correlation length that diverges \cite{Sachdev2011,Tsuda2013}. Passing through the phase transition point results in the ground state of a system being restructured, which is manifested by a certain non-analiticity in the ground state energy of a quantum Hamiltonian. It is therefore not surprising that there exists a final overlap between two different ground states of the system, which is regarded as a meaningful source of information on the quantum phases of a system and can be rigorously worked out within the fidelity approach \cite{Venuti2007,Damski2016}.

\subsection{Sampling spin configurations}

\begin{figure}[ht]
\centering
\vspace{-10mm}
\includegraphics[width=0.5\textwidth]{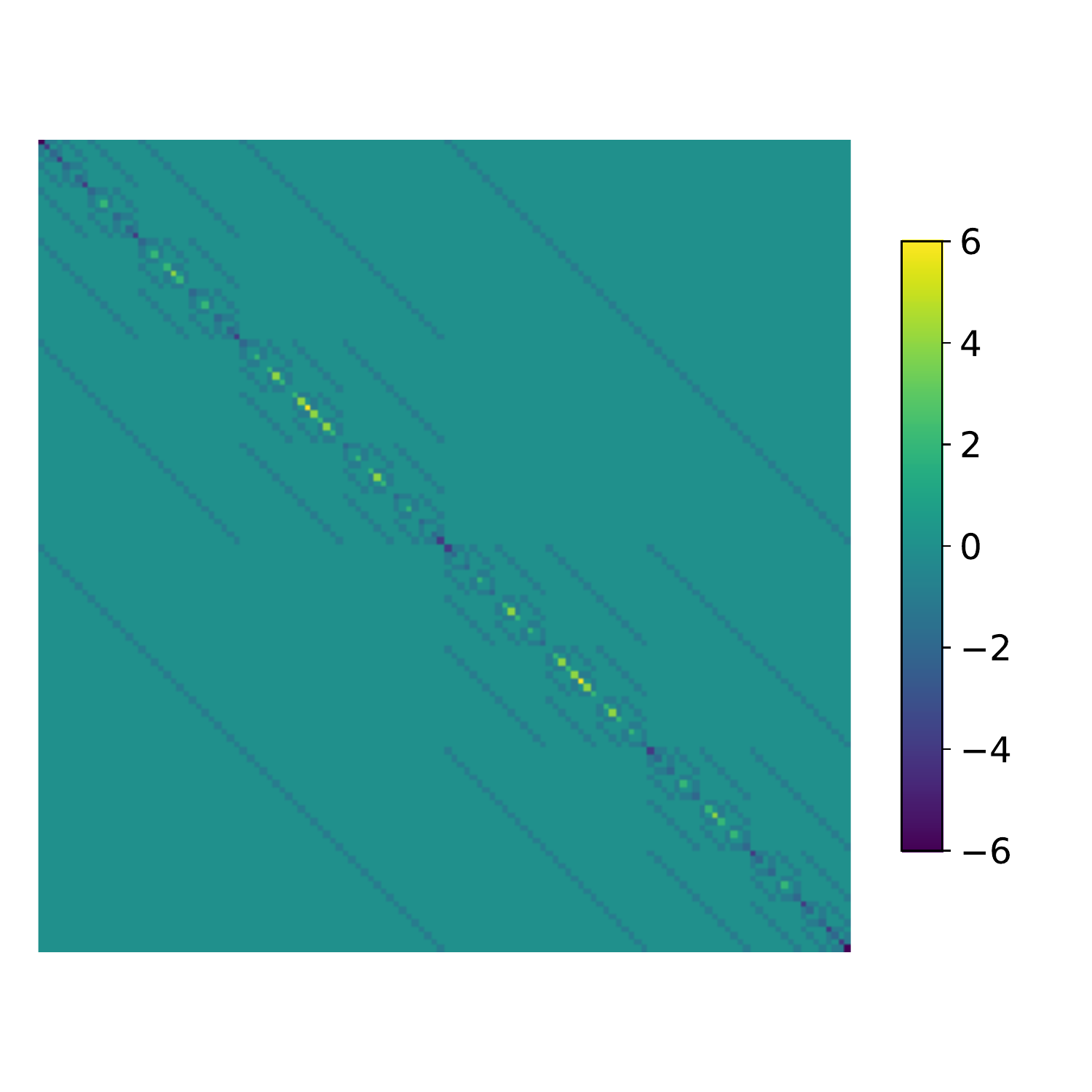}
\vspace{-10mm}
\caption{Heatmap of the matrix that corresponds to a one-dimensional quantum TFIM with the Hamiltonian \eref{eq:1} and $L=7$ spins at criticality $\tau/J=1$ in computational basis.}
\label{fig:heatmap}
\end{figure}

In this section, we briefly describe the sampling routine we used for the interacting spin models, described by the Hamiltonians \eref{eq:1} and \eref{eq:2}. Note that the Hamiltonians \eref{eq:1} and \eref{eq:2} are sparse in the standard basis matrices with most of the elements being zero, as schematically shown in \Fref{fig:heatmap} for a system of $L=7$ spins.

For small systems the exact diagonalization of the Hamiltonians of Eqs.~\eref{eq:1} and \eref{eq:2} is possible. Let a $2^L$-dimensional vector $\vert g\rangle$ be the ground state of this system. In the computational basis the vector
\begin{equation}
    \vert g\rangle=\sum\limits_{i_1,i_2,\ldots,i_L=\uparrow,\downarrow}\alpha_{i_1i_2\ldots i_L}\vert i_1\rangle\vert i_2\rangle\ldots\vert i_L\rangle,
\end{equation}
is purely determined by $2^L$ complex-valued decomposition components $\alpha_{i_1i_2\ldots i_L}$ in the basis $\vert i_k\rangle=\{\vert\uparrow\,\rangle,\vert\downarrow\,\rangle\}$, with $k=1,\ldots,L$, which are known to give the probability distribution $p_{i_1i_2\ldots i_L}=|\alpha_{i_1i_2\ldots i_L}|^2$ of a particular spin configuration $\vert i_1\rangle\vert i_2\rangle\ldots\vert i_L\rangle$, which we refer to as a {\it bitstring} and later represent explicitly as strings of $0$'s and $1$'s. Thus, sampling the physical system specified by the Hamiltonian \eref{eq:1} might be approached by sampling each bitstring with the corresponding probabilities $p_{i_1i_2\ldots i_L}$.

\subsection{Neural network architecture}

\begin{figure}[h]
\centering
\includegraphics[width=0.6\textwidth]{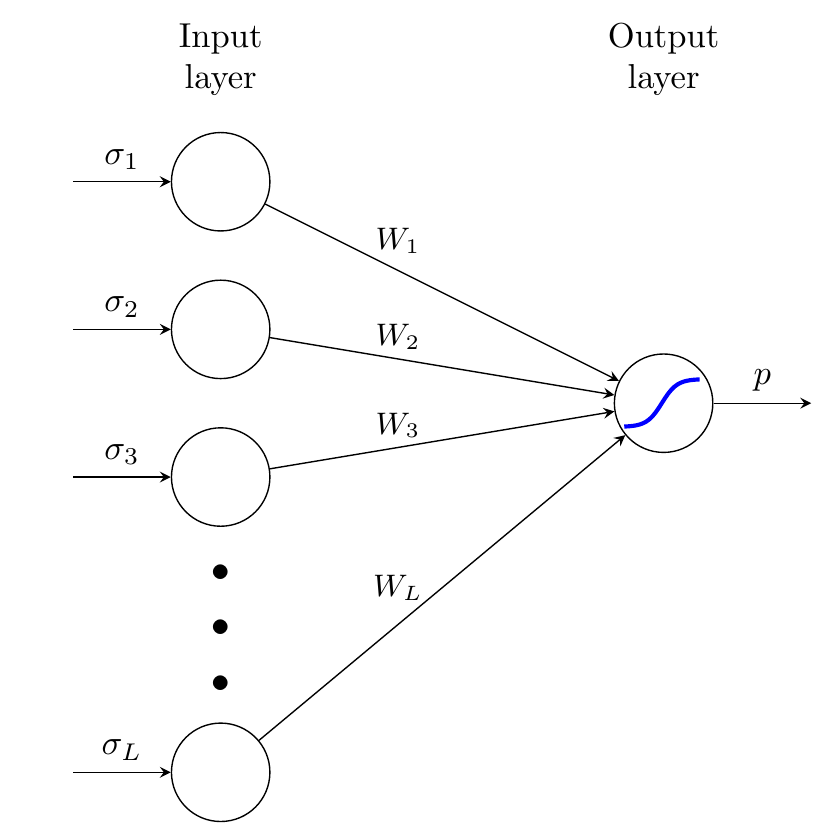}
\caption{The neural network design. $W_i$ denotes the weights connecting the input layer neurons with the output neuron, $\sigma_i$ denotes a spin value in the $z$-basis fed into the input layer, the solid blue line denotes the sigmoid activation function which for the output neuron.}
\label{fig:nn}
\end{figure}

We use a neural network architecture that consists of an input layer and one output neuron, corresponding to a binary classifier. The sampled bitstrings serve as input data. Noteworthy, any hidden layers are absent. The output is prescribed to take value $0$ when an input spin configuration is drawn from the ground state prescribed by $\tau_1=0.01$ ($\gamma_1 = -1$), whereas if the configuration is taken from $\tau_i$ ($\gamma_i$), the neuron is prescribed to take the value $1$. We also discuss results of numerical simulations with other starting points, $\tau$ and $\gamma$---see Section 4. The neural network architecture used is shown in \Fref{fig:nn}.

The linear combination of the spins' $z$-projections $\{\sigma_i\}$ is fed into the neural network via the input layer, followed by a nonlinear activation of the output neuron
\begin{equation}
    p(\{\sigma_i\}) = \mathrm{sigm}\left(\sum_{i=1}^L W_i \sigma_i + b_i \right),
\end{equation}
with $\mathrm{sigm}(x) := (1+e^x)^{-1}$ being the sigmoid function and the binary cross-entropy
\begin{equation}
    H(p)= - \sum_{i=1}^{N_{\tiny{\textrm{train}}}}\{y_i \cdot \log  p\left( y_i \right) + \left( 1-y_i \right) \cdot \log \left[1-p \left( y_i \right) \right]\},
\end{equation}
serving as the loss-function. Such a simple form of the neural-network architecture results in high computational speed. The neural network outcome is the probability that the input state should be classified as belonging to the respective ground state specified by the control parameter value. Here, for a set of training data points $\{\sigma_i\}$ with $1\leq i\leq N_\mathrm{train}$ the neural network predicts the probability $p(y_i)$ for labels $y_i\in\{0,1\}$. We make use of two labels, ``Phase 1'' and ``Phase 2'', namely magnetically ordered and disordered phases for TFIM as well as X- and Y-ordered phases for the anisotropic XY model depending on parameters $\tau$ and $\gamma$ respectively. While the parameters of the neural network, the weights and the biases, are updated using the {\it RMSProp} algorithm \cite{hinton2012neural}.

\subsection{The Algorithm}

In our numerical simulations, for chains of $L=20$ spins we explore the model described by \Eref{eq:1} throughout the region $0.01J\leq\tau\leq2J$ with $D=40$ steps, $\tau=\{\tau_i\}_{i=1}^D$ and $N=10^4$ spin configurations to be sampled for each value of $\tau_i$. Afterwards, a feed-forward neural network $\mathbb{N}_i$ is trained to classify the bitstrings sampled for $\tau_1=0.01$ from those for $\tau_i$ with $i>1$. Finally, we end up with $D-1$ pairs of $(P_i,\tau_i)$ with $P_i \in (0,1)$ being the mean output of the neural network evaluated on the samples drawn from the probability distribution given by the ground state of $H(\tau_i)$. In what follows, we show that the value of $P$ with respect to $\tau$ dramatically changes signalling a phase transition. We apply a similar procedure to the anisotropic XY model with the anisotropy parameter $-1 \leq \gamma \leq 1$ starting with $\gamma_1 = -0.99$. The result is then averaged over 40 runs to rid possible effects caused by random initialization of the neural networks' parameters (displayed as shadows in the plots). The results described in the paper are obtained under the following conditions: we divide our initial dataset of $10^4$ configurations per control parameter value into two subsets, namely training and testing set, which is in line with commonly accepted ratio of 80\% for training, 20\% for testing. As for the performance of the algorithm, generally, for the binary classification problem, the performance measure would be the accuracy of classification, which reached at 95\% for the training set and 84\% for the test set within all numerical experiments conducted.

\section{Results}

\begin{figure}[ht]
\centering
\includegraphics[width=0.6\textwidth]{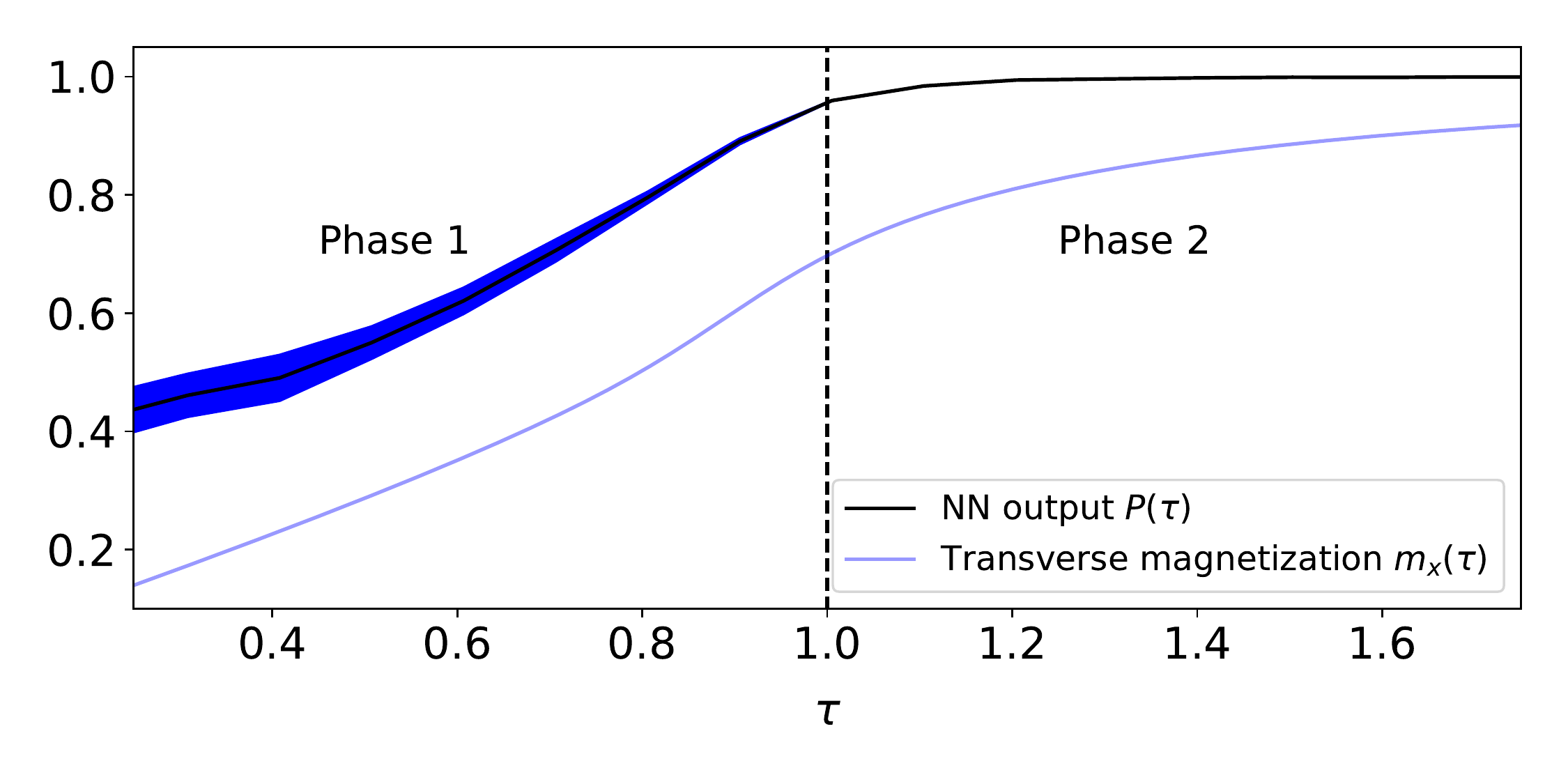}
\caption{The output of trained neural networks as a function of the transverse magnetic field $\tau$, for $L=20$ spins on a TFIM chain with open boundary conditions, qualitatively reproduces the behavior of transverse magnetization as obtained by exact solution to TFIM.}
\label{fig:tfim_output}
\end{figure}

Below, we present and discuss the results of our numerical simulations, demonstrating how the neural network architecture and the corresponding algorithm described in Section 3 are capable of probing the phase crossover point for the described models. In \Fref{fig:tfim_output}, we show how our setup performs for TFIM on an open chain of $L=20$ spins together with the transverse magnetization defined as
\begin{equation}
    m_x = \frac{1}{L} \sum_{i=1}^{L} \langle \sigma_i^x \rangle,
\end{equation}
where averaging is done over the ground state. As expected, the neural network learns the order parameter due to the linearity of the latter as a function of spin projections. Note however, that while the resulting curve is remarkably close to the transverse magnetization curve, there was no information about the $x$-projections of the spin measurements in our setup, but only the measurements in the $z$-basis.

\begin{figure}[h]
\centering
\includegraphics[width=0.6\textwidth]{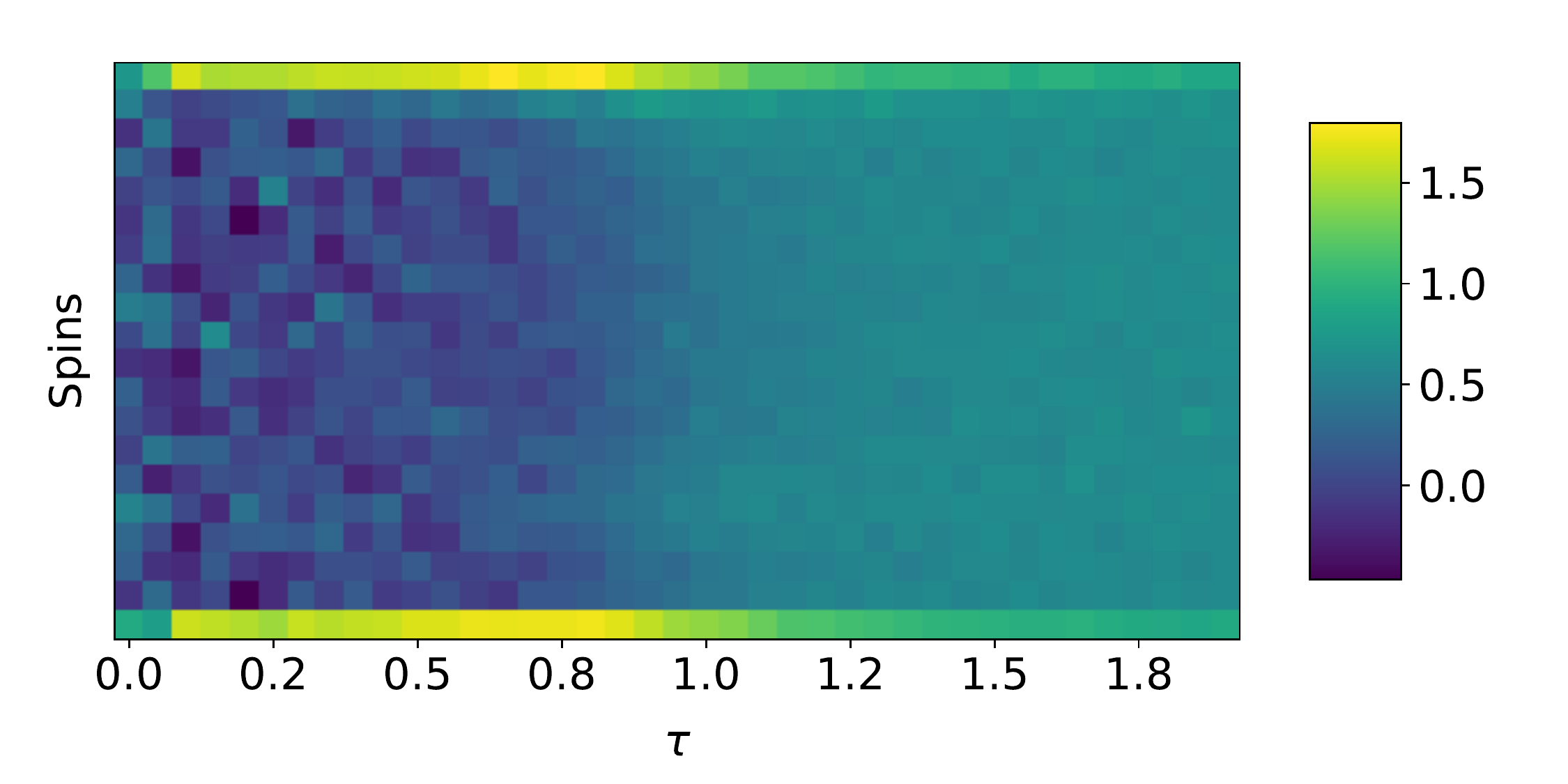}
\caption{Heatmap of the weights $W_i$ of the neural networks for a TFIM chain of $L=20$ spins with open boundary conditions depending of the magnetic field strength.}
\label{fig:tfim_weights}
\end{figure}

Unlike in previous studies, for example \cite{Arai2018}, the simplicity of a neural network used for the simulations makes direct visualization of the weights straightforward owing to their vectorial nature. \Fref{fig:tfim_weights} clearly displays the crossover in the neighborhood of criticality, making these results intuitively clear and interpretable in contrast to usual deep learning routines \cite{Montavon2018,Liu2019}. Each vertical row in \Fref{fig:tfim_weights} corresponds to a set of coefficients $z$-components of spins are multiplied by before transferring the whole sum to the activation function of the output neuron. Thus, the model actually mimics $z$-projections of spin configurations given the transverse magnetic field value $\tau$. The latter explains why the rows in the heatmap are uniform in the ferromagnetic limit and take random values in the disordered phase. Note that the boundary coefficients are different because of the open boundary conditions.

\begin{figure}[ht]
\centering
\includegraphics[width=0.6\textwidth]{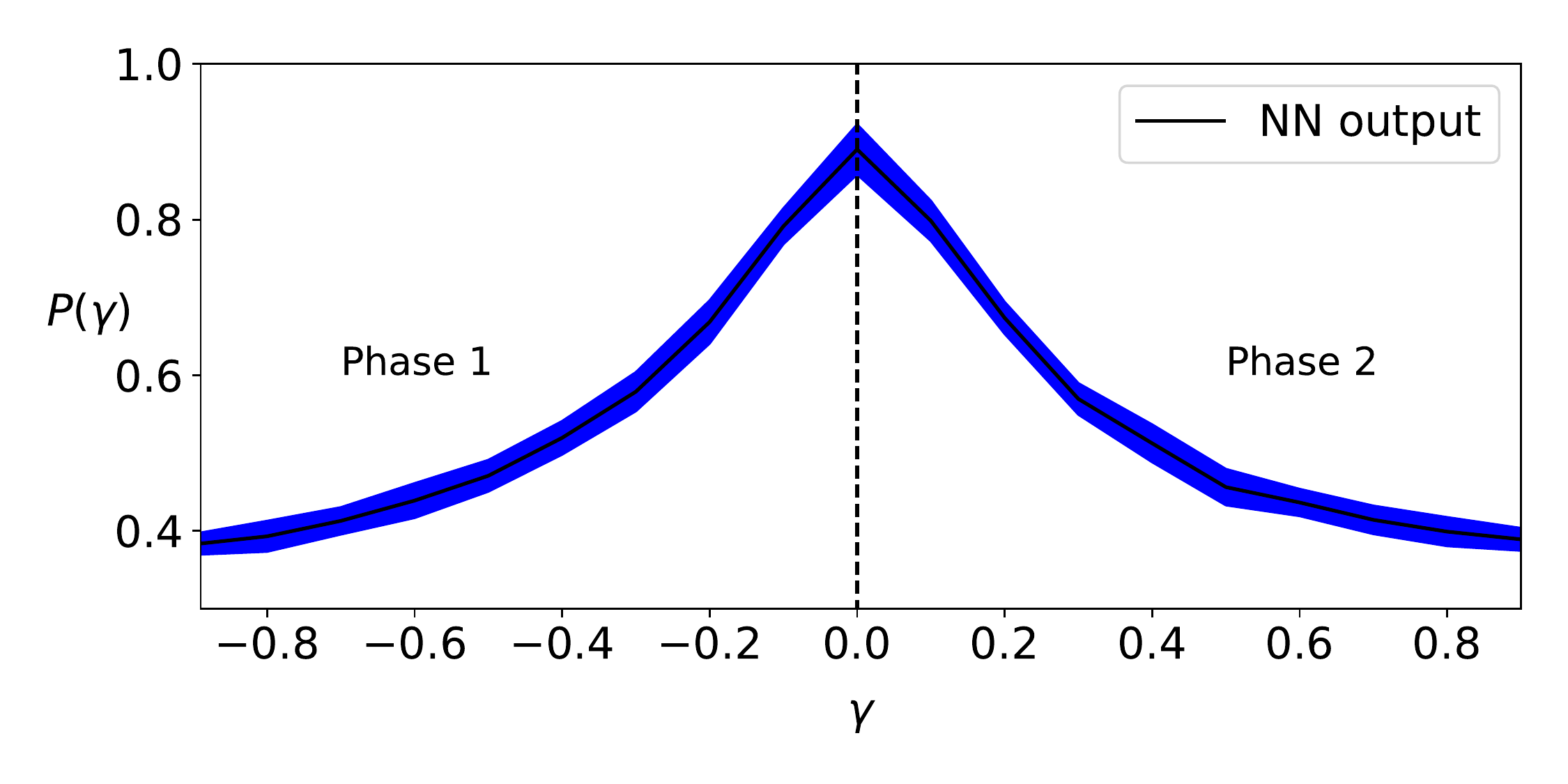}
\caption{The output of trained neural networks as a function of the anisotropy parameter $\gamma$ for $L=20$ spins on an anisotropic XY chain with open boundary conditions.}
\label{fig:xy_output}
\end{figure}

In \Fref{fig:xy_output}, we show the result for an anisotropic XY chain of $L=20$ spins. In this plot, one can clearly see the phase crossover induced by the change of $\gamma$ which is a sign of a well-studied anisotropy-induced phase transition in an infinite system \cite{quan2009finite}, similarly to the phase transition induced by the critical value of the magnetic field. Again, while our algorithm is given information about the $z$-components of spins, it is capable of exposing a phase crossover induced by the anisotropy in the $x$-$y$ plain. In this case, there is no direct correspondence to any observable.

\begin{figure}[ht]
\centering
\includegraphics[width=0.6\textwidth]{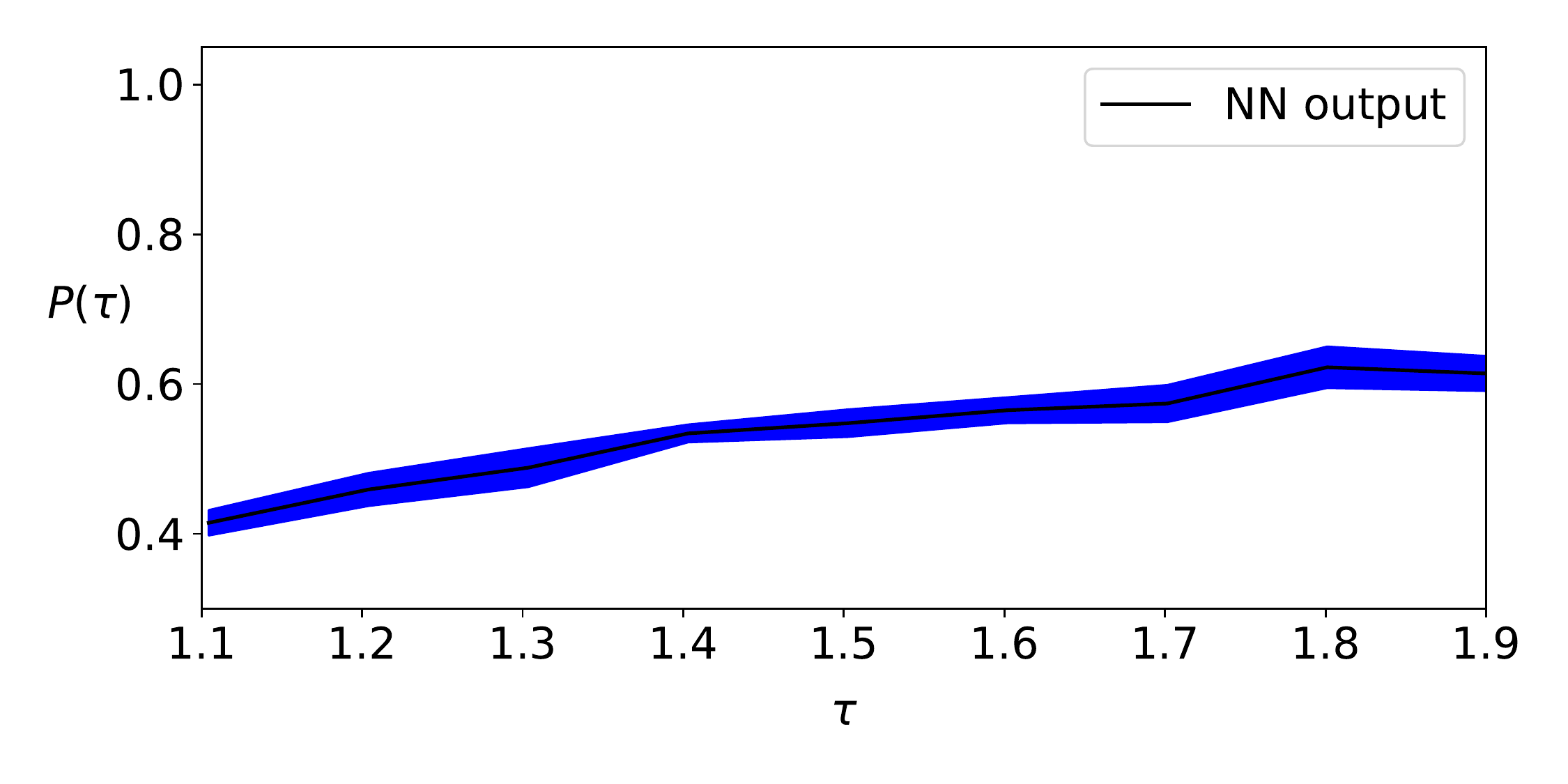}
\caption{The output of trained neural networks as a function of the transverse magnetic field $\tau$, for $L=20$ spins on a TFIM chain with open boundary conditions, on condition that $\tau_0=1.0$ and $\tau_{D}=2.0$.}
\label{fig:tfim_output_same_phase_scan}
\end{figure}

\begin{figure}[ht]
\centering
\includegraphics[width=0.6\textwidth]{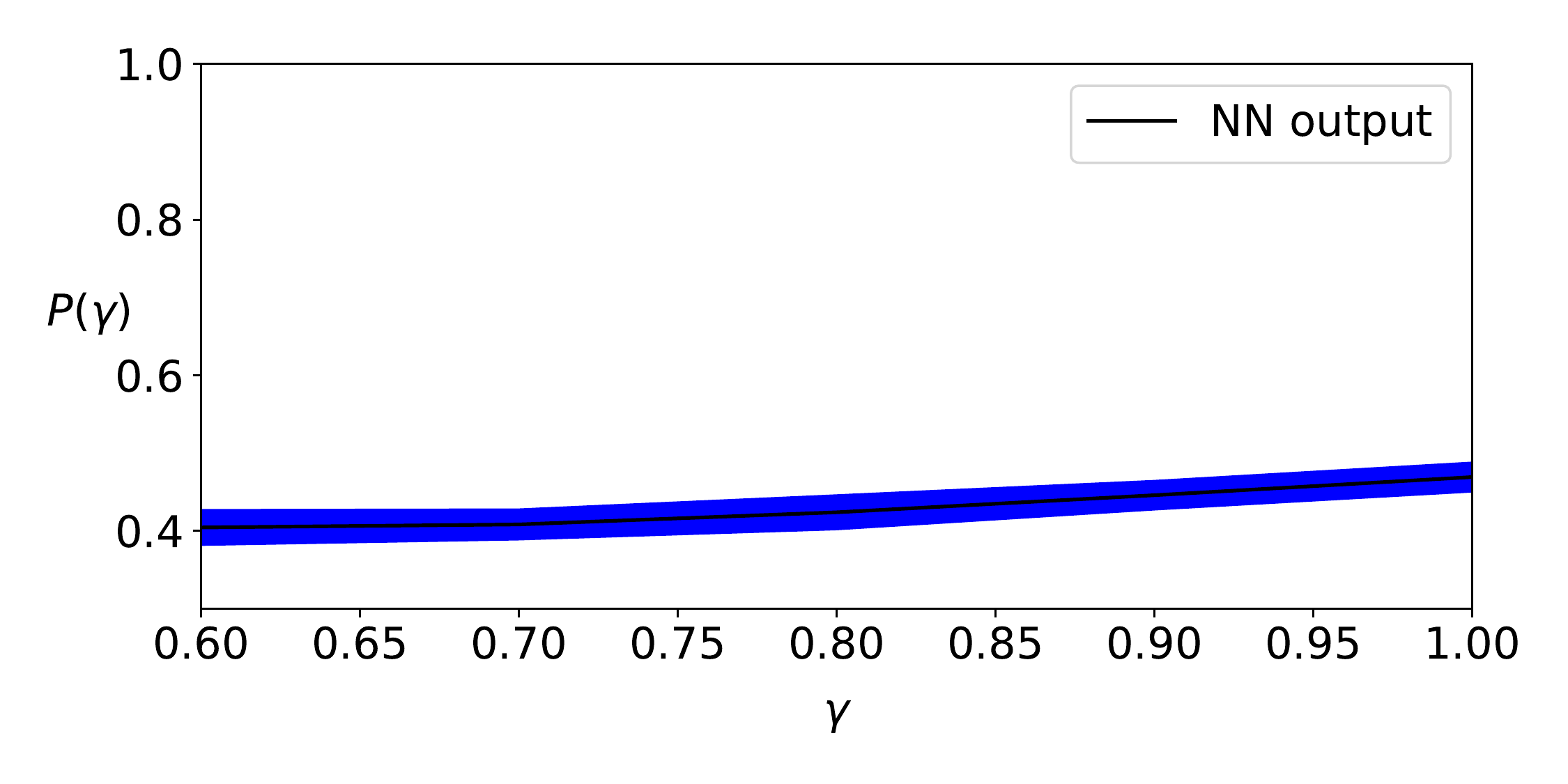}
\caption{The output of trained neural networks as a function of the anisotropy parameter $\gamma$, for $L=20$ spins on an anisotropic XY chain with open boundary conditions, provided that $\gamma_0=0.5$ and $\gamma_{D}=1.0$.}
\label{fig:xy_output_same_phase_scan}
\end{figure}

To further validate the proposed algorithm we provide outcomes of the neural network(s) for different starting reference points---that is, other than $\tau_0=0.01$ or $\gamma_0=-1$ as discussed earlier.  We hence perform the scan as described in Section 3 while staying potentially in the same phase. The results of our numerical findings for $\tau_0=1$ and $\gamma_0=0.6$ are shown in \Fref{fig:tfim_output_same_phase_scan} and \Fref{fig:xy_output_same_phase_scan}.

\begin{figure}[ht]
\centering
\includegraphics[width=0.6\textwidth]{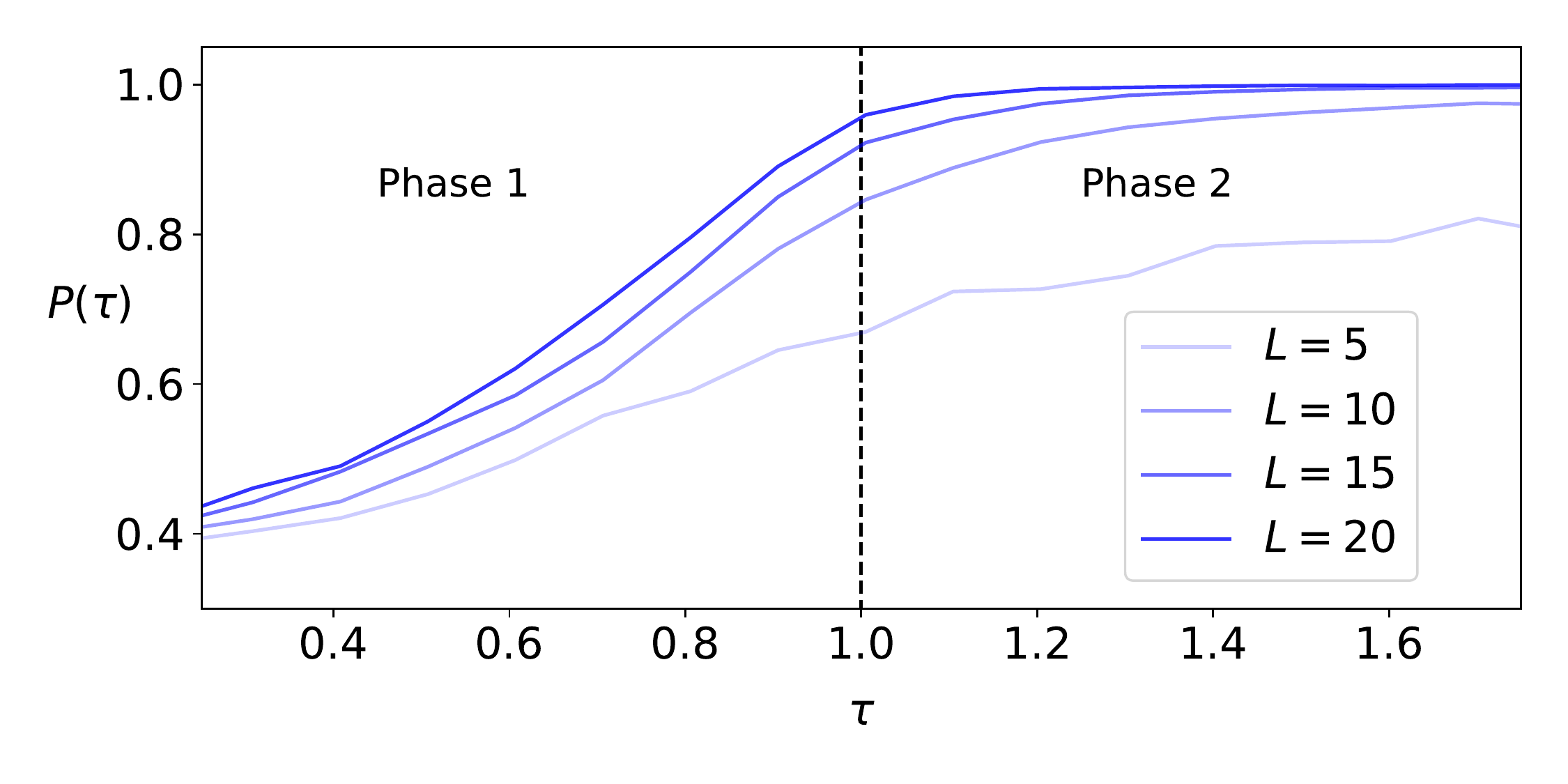}
\caption{The output of trained neural networks as a function of the transverse magnetic field $\tau$, for $L=20,15,10,5$ spins on a TFIM chain with open boundary conditions.}
\label{fig:tfim_output_scaling}
\end{figure}

\begin{figure}[ht]
\centering
\includegraphics[width=0.6\textwidth]{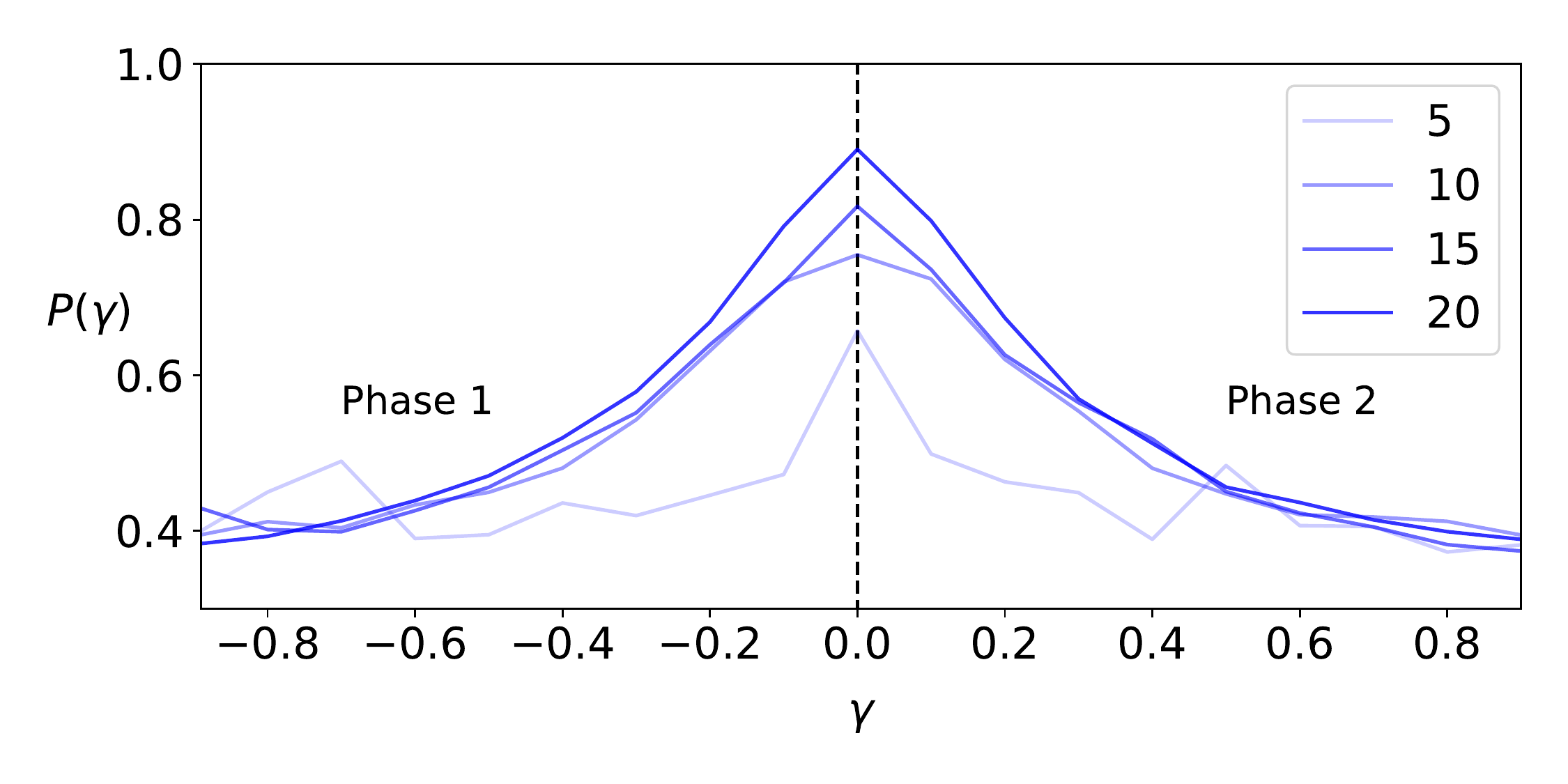}
\caption{The output of trained neural networks as a function of the anisotropy parameter $\gamma$ for $L=20,15,10,5$ spins on an anisotropic XY chain with open boundary conditions.}
\label{fig:xy_output_scaling}
\end{figure}

To address finite-size scaling on the performance of our algorithm we provide the results of numerical simulations for phase classification depending on the number of spins $L$ in \Fref{fig:tfim_output_scaling} and \Fref{fig:xy_output_scaling}. These findings suggest that classification robustness increases with the number of spins, while starting from $L=20$ allowed us to correctly separate phases.

\section{Conclusion}

In this paper, we have considered the simplest neural network architecture with no hidden layers present and applied it to study the finite-size phase crossovers in the quantum transverse field Ising model and the quantum anisotropic XY model on a one-dimensional chain. We were able to distinguish the regions of different phases using neural networks {\it without} prior knowledge of the phase diagram by observing the corresponding phase boundary crossover in a finite-size system. Relative simplicity of the machine learning setup allowed us to visualize the weights of the corresponding neural network and unambiguously relate this plot to configuration of different spin orderings.

\section{Data availability}

The data that support the findings of this study are available upon reasonable request from the authors.

\section{Acknowledgements}

The authors are thankful to Anastasiia Pervishko and Sebastian Wetzel for fruitful discussions. ZZ acknowledges support from the J\'anos Bolyai Research Scholarship, the UKNP Bolyai+ Grant, and the NKFIH Grants No. K124152, K124176 KH129601, K120569, and from the Hungarian Quantum Technology National Excellence Program, Project No.~2017-1.2.1-NKP-2017-00001. The work of DY was supported by the Russian Foundation for Basic Research Project No.~19-31-90159. JB acknowledges support from the research project, Leading Research Center on Quantum Computing (Agreement No.~014/20).
\section{References}

\bibliographystyle{unsrt}
\bibliography{main.bbl}

\end{document}